\documentclass{icrc2009}

\usepackage{graphicx}   % for including figures
\usepackage[caption=false]{caption}    % for captions
\usepackage[font=footnotesize]{subfig} % subfig.sty for a double column floating figure using two subfigures
\usepackage{fixltx2e}
\usepackage{url}

\newcommand{\shorttitle}[1]%
{\markboth{Proceedings of the 31\MakeLowercase{$^{st}$} ICRC, {\L}\'{o}d\'{z} 2009}{#1} }
\def\simleq{\; \raise0.3ex\hbox{$<$\kern-0.75em \raise-1.1ex\hbox{$\sim$}}\; }
\def\simgeq{\; \raise0.3ex\hbox{$>$\kern-0.75em \raise-1.1ex\hbox{$\sim$}}\; }

\newcommand{\GeV}{{\rm GeV}}

\newcommand{\kpc}{{\rm kpc}}

\begin{document}
\title{Possible Interpretations of the High Energy Cosmic Ray Electron Spectrum measured with the Fermi Space Telescope}

\author{\IEEEauthorblockN{D.~Grasso \IEEEauthorrefmark{1} on behalf of the Fermi collaboration}\\
\IEEEauthorblockA{\IEEEauthorrefmark{1} Istituto Nazionale di Fisica Nucleare, Sezione di Pisa,  I-56127 Pisa, Italy}}

\shorttitle{Fermi Collaboration - Interpretations of the high energy electron spectrum}
\maketitle

\begin{abstract}
The Fermi Large Area Telescope has provided the measurement of the high energy (20 GeV to 1 TeV) cosmic ray electrons and positrons spectrum with unprecedented accuracy. This measurement represents a unique probe for studying the origin and diffusive propagation of cosmic rays as well as for looking for possible evidences of Dark Matter. In this 
contribution we focus mainly on astrophysical sources of cosmic ray electrons and positrons which include the standard primary and secondary diffuse galactic contribution, as well as nearby point-sources which are expected to contribute more significantly to higher energies. In this framework, we discuss possible interpretations of Fermi results in relation with other recent experimental data on energetic electrons and positrons (specifically the most recent ones reported by PAMELA, ATIC, PPB-BETS and H.E.S.S.).
\end{abstract}
%\maketitle

\begin{IEEEkeywords}
Cosmic ray electrons, Fermi Gamma Ray Telescope
\end{IEEEkeywords}

%\linenumbers

\section{Introduction}

Prior to 2008, the high energy electron spectrum was measured by balloon-born experiments \cite{Kobayashi:2003kp} and by a single space mission AMS-01 \cite{ams1}. Those data are compatible with a featureless power law spectrum within their errors. 
This is in agreement with theoretical predictions (for a recent review see \cite{Strong:2007nh}) assuming: i) that the source term of CR electrons is
treated as a time-independent and smooth function of the position in the Galaxy, and the energy dependence is assumed to be a power law; ii) that the propagation is described by a diffusion-loss equation whose effect is to steepen the spectral slope respect to the injection. 
Possible deviations from a simple power law spectrum may, however, be expected above several hundred GeV as a consequence of synchrotron radiation and 
Inverse Compton (IC) energy losses which, at those high energies, limit the electron propagation length to a distance comparable to the mean distance between 
astrophysical sources \cite{Aharonian:95,Pohl:1998ug} or because the possible presence of exotic sources.    

Few months ago, the ATIC balloon experiment  \cite{atic} found a prominent spectral feature  at around 600 GeV in the total electron spectrum.  Furthermore, the H.E.S.S. \cite{hess,hess:09} atmospheric Cherenkov telescope reported a significant steepening of the electron plus diffuse photon spectrum above 600 GeV. Another independent indication of the presence of a possible deviation from the standard picture came from the recent measurements of 
the positron to electron fraction, e$^+$ /(e$^-$+ e$^+$), between 1.5 and 100 GeV by the PAMELA satellite experiment \cite{PAMELA,PAMELA_Nature}.  PAMELA found that the positron fraction changes slope at around 10 GeV and begins to increase steadily up to 100 GeV. 
This behavior is very different from that predicted for secondary positrons produced in the collision of CR nuclides with the interstellar medium (ISM).

Recently the experimental information available on the CRE spectrum has been drastically expanded as the Fermi Collaboration has reported a high precision measurement of the electron spectrum from 20 GeV to 1 TeV performed with its Large Area Telescope (LAT) 
 \cite{Fermi_CRE_1}.  A simple power law fit of the Fermi-LAT electron energy spectrum (see Fig.\ref{fig:elepos_242reac}) is possible giving: 
%\begin{equation}
$\displaystyle J_{e^\pm} = (175.40 \pm 6.09) \left(\frac{E}{1~\GeV} \right)^{-(3.045 \pm 0.008)}  \GeV^{-1} {\rm m}^{-2} {\rm s}^{-1} {\rm sr}^{-1} $                        
%\label{eqn:eqn1}
%\end{equation}
with $\chi^2$ = 9.7 (for 23 d.o.f.) where statistical and systematic (dominant) errors have been, conservatively, added in quadrature.
The electron spectrum measured by Fermi-LAT reveals a hardening at around 70 GeV and a steepening above $\sim 500$ GeV.
Although the significance of those features is low within current systematics, they suggest the presence of more components in the electron 
high energy spectrum.  It is also worth noticing here that the hard electron spectrum observed by this experiment exacerbates the discrepancy between the predictions of standard CR theoretical models and the positron faction excess measured, most conclusively, by PAMELA \cite{PAMELA,PAMELA_Nature}. 
This makes the exploration of some non-standard interpretations more compelling. 

\section{Conventional interpretation}\label{sec:GCRE}

\begin{figure}[!t]
 \centering
 \includegraphics[width=3in]{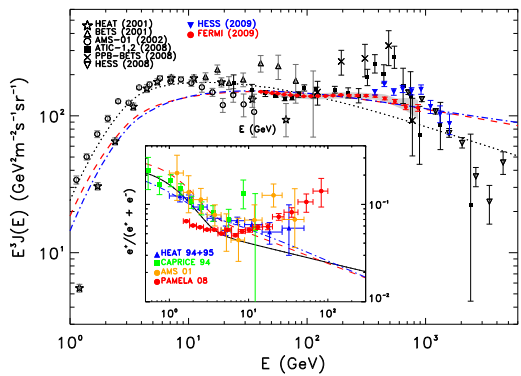}
 \caption{Fermi-LAT CRE data  \cite{Fermi_CRE_1}, as well as several other experimental data sets,
 are compared to the $e^- + e^+$ spectrum modeled with GALPROP.
 The gray band represents systematic errors on the CRE spectrum measured by Fermi-LAT.
 The dotted (black) line correspond to the conventional model used in \cite{Strong:04} to fit   
 pre-Fermi data conventional model. 
 The dashed (red) and dash-dotter (blue) lines are obtained with modified injection indexes $\gamma_0 = 2.42$  
 (for $\delta = 0.33$) and $\gamma_0 = 2.33$ (for $\delta = 0.6$) respectively.
 In the insert  the positron fraction for the same models is compared with experimental data.
 All models account for solar modulation in the force field approximation assuming a potential $\Phi = 0.55~{\rm GV}$.     
 }
 \label{fig:elepos_242reac}
\end{figure}

We start considering a possible interpretation of Fermi-LAT CRE data in terms of a conventional {\it Galactic CR electron scenario} 
(GCRE) model assuming that electrons sources are continuously distributed in the Galactic disk and that  positrons are only produced by the collision of primary CR nuclides with the interstellar gas. 
To this purpose we use the GALPROP numerical CR propagation code \cite{Moskalenko:01}.
We consider here two reference conventional models with injection spectral index $\gamma_0 = 2.42$ above 4 GeV,
if the value of power law index of the diffusion coefficient dependence on energy is $\delta = 0.33$, 
and  $\gamma_0 = 2.33$ if $\delta = 0.6$ (see Tab. 1 in \cite{interpretation_paper} for more details about those models). As shown in Fig. \ref{fig:elepos_242reac} 
these models provide a good representation of Fermi-LAT CRE data.
In the same figure we also show for comparison a conventional model with $\gamma_0 = 2.54$ which was already successfully used to interpret  
pre-Fermi CRE data \cite{Strong:04} and the diffuse gamma-ray emission measured by Fermi-LAT at intermediate Galactic latitudes \cite{GeVexc}. 

GCRE models, however,  face a series of problems, when compared with other experimental data sets, namely:
i) they display a significant tension with respect to low energy pre-Fermi data,  AMS-01\cite{ams1} and HEAT \cite{DuVernois:2001bb} most noticeably;
ii) they exceed H.E.S.S. data above 1 TeV; 
iii)  most seriously,  the positron fraction  $e^+/(e^+ + e^-)$ they predict is not consistent with that measered by PAMELA \cite{PAMELA,PAMELA_Nature}
(see the insert in Fig. \ref{fig:elepos_242reac}) .
While item (ii) may be interpreted as a consequence of  the stochastic nature of astrophysical sources (see Sec. 2.2 in \cite{interpretation_paper} and ref.s therein) the other caveats are most serious.
For these reasons in the following sections we consider the possibility that an additional electron and positron primary component contribute to the observed Fermi-LAT, H.E.S.S. and PAMELA data at high energy.  

\section{Pulsar interpretation}\label{sec:pulsars}

Pulsars are undisputed sources of relativistic electrons and positrons, believed to be produced in their magnetosphere and subsequently possibly reaccelerated by the pulsar wind or in the supernova remnant shocks (see e.g. \cite{Harding:87,Zhang:01}).  
 For bright young pulsars the maximal acceleration energy can be as large as $10^3$ TeV. This quantity decreases for middle-age or, so called, {\it mature}
 pulsars (i.e. with age $10^4 \simleq T \simleq 10^6~{\rm yr}$ ). Electrons and positrons are expected to be liberated into the ISM only after pulsar wind nebulae or the surrounding supernova remnant merge into the ISM,  $10^{4}$ - $10^{5}$ years after the pulsar birth. This process should be relatively fast so that  {\it mature} pulsars can effectively be treated as burst-like sources of electrons and positrons.  
 The possible role of these source explaining the PAMELA positron fraction anomaly \cite{PAMELA,PAMELA_Nature} has been discussed in several papers 
 (see e.g. \cite{Zhang:01,Hooper:2008kg,Profumo:2008ms} and ref.s therein).

We compute the spectrum of electrons and positrons from each pulsar by following the approach reported in the appendix of \cite{interpretation_paper}.
The basic input is the  $e^\pm$ energy release of each mature pulsar that we determine by integrating the observed spin-down luminosity over time giving 
 (see e.g. \cite{Profumo:2008ms}) $E_{e^\pm}  \simeq \eta_{e^\pm}~  {\dot  E}_{\rm PSD}~\frac{T^2}{\tau_0}$
where $ {\dot  E}_{\rm PSD}$ is the present time spin-down luminosity determined form the observed pulsar timing, $T = P/2{\dot P}$ (where $P$  is the pulsar period) 
the pulsar age,  and $\eta_{e^\pm}$ is the $e^\pm$ pair conversion efficiency of the radiated electro-magnetic energy. For the characteristic luminosity decay time we assume 
$\tau_0 = 10^4~{\rm years}$ as conventionally adopted for mature pulsars. 
The setup we use here to model the large-scale GCRE spectrum is a slightly rescaled version of  the conventional model used to interpret pre-Fermi data \cite{Strong:04}
 (we reduced the electron flux normalization by a factor $\sim 0.95$ respect to that model so to leave room to the extra pulsar $e^\pm$ component). 

\begin{figure}[!t]
  \centering
  \includegraphics[width=3in]{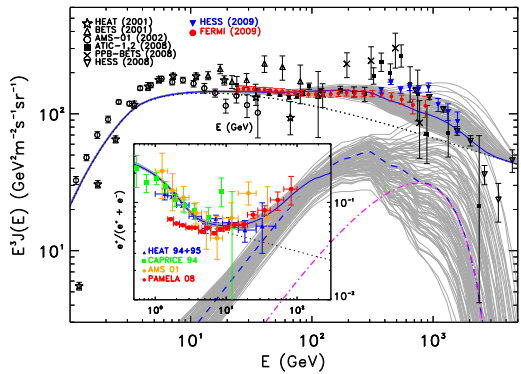} 
  \caption{The $e^- + e^+$ spectrum from pulsars plus the Galactic (GCRE) component with experimental 
  data (dotted line). Each gray line represents the sum of all pulsars for a particular combination of pulsar parameters.  
  The dashed (pulsars only) and solid (pulsars + GCRE component) blue lines
   correspond to a representative choice among that set of possible realizations.  
  The dot-dashed (purple) line represents the contribution of Monogem pulsar in that particular case.
  Note that for graphical reasons here Fermi-LAT statistical and systematic errors are added in quadrature.
  In the insert the positron fraction for the same models is compared with experimental data.
  Solar modulation is accounted as done in Fig.\ref{fig:elepos_242reac}.
  }
  \label{fig:elepos_random_pulsars}
\end{figure}

In general several pulsars contribute to the electron and positron fluxes reaching the Earth.  For this reason we summed the contribution to the electron and positron 
flux of all pulsars in ATNF radio pulsar catalogue (http://www.atnf.csiro.au/research/pulsar/psrcat/ )  \cite{Manchester:05} with distance $d < 3~\kpc$  and age 
$T > 5 \times 10^4~{\rm yr}$ ( $\sim 150$  pulsars). More distant pulsars give a negligible contribution at the energies considered here;
we assume that electron accelerated in younger pulsars are still confined in their nebulae (lowering this limiting age would not change significantly our results).
For each of these pulsars we use the spin-down luminosity given in the catalogue and randomly vary the relevant parameter in the following representative ranges:
$800 < E_{\rm cut} < 1400~\GeV$,  $10 < \eta_{e^\pm} <  30~\%$  and $5 <  (\Delta t / 10^4~{\rm yr} ) <  10$ and $1.5 < \Gamma < 1.9$.
These ranges of parameter are compatible with our observational and theoretical knowledge of particle acceleration in PWNe (see e.g. \cite{Aharonian_book}).
Following this approach we find that  Fermi-LAT CRE data comfortably lie within the bands of those realizations (see Fig.~\ref{fig:elepos_random_pulsars}) and are 
in reasonable agreement with the positron fraction measured by PAMELA (see the insert in the same figure).  
It should be noted that the ATFN catalogue does not include all pulsars. Some pulsars radio beams are not pointing toward us and also selection effects in the radio detection intervene to reduce the number of the observed  pulsars. Furthermore, the recent discovery of a population of radio-quiet gamma-ray pulsars by Fermi-LAT \cite{blind_search} has demonstrated that  those pulsars are a significant fraction of the total pulsar set. 
 We do not expect, however, that the average spectral shape would change significantly by accounting for pulsars not included in the ATFN catalogue.
The larger electron and positron primary flux due to the contribution of those sources can be compensated by invoking a smaller pair conversion efficiency 
$\eta_{e^\pm}$ making this scenario even more appealing. While selection effects may lead to underestimate older pulsar at large distance, their
role is almost negligible at the energies of interest here.

\section{Dark matter interpretation}\label{sec:DM}

Here we briefly discuss about the alternative possibility of interpreting in Fermi-LAT CRE data in terms of an electron and positron component originated from the pair-annihilation of Galactic dark matter (DM).
The new Fermi-LAT data affect a dark matter interpretation of CRE data in at least three ways:
%\begin{enumerate} 
{\it i)} The rationale to postulate a particle dark matter mass in the 0.5 to 1 TeV range, previously motivated by the ATIC data and the detected ``bump'', is now much weaker, if at all existent, with the high statistics Fermi-LAT data; 
{\it ii)}  CRE data can be used, in the context of particle dark matter model building, to set constraints on the pair annihilation rate or on the decay rate, for a given dark matter mass, diffusion setup and Galactic halo model;
{\it iii)}  as discussed in Sec.\ref{sec:GCRE}, unlike the Fermi-LAT CRE result, the PAMELA positron fraction measurement requires one or more additional primary sources in addition to the standard GCRE component, as discussed in Sec.~\ref{sec:GCRE}; if the PAMELA data are interpreted in the context of a dark-matter related scenario, Fermi-LAT data provide a correlated constraint to the resulting total CRE flux.

Here we consider the following representative class of models:
\begin{enumerate}
\item {\em Pure $e^\pm$ models}: for this class of models, the dark matter pair annihilation always yields a pair of monochromatic $e^\pm$, with injection energies equal to the mass of the annihilating dark matter particle. Such models arise for instance in the context of frameworks where the dark matter sector is {\em secluded} 
\cite{Pospelov:2007mp}, and the dark matter pair-annihilates into a {\em light gauge boson} which can then kinematically decay only into $e^\pm$ 
\cite{ArkaniHamed:2008qn}.
\item {\em Lepto-philic models}: here we assume a democratic dark matter pair-annihilation branching ratio into each charged lepton species: 1/3 into $e^\pm$, 1/3 into $\mu^\pm$ and 1/3 into $\tau^\pm$. Here too antiprotons are not produced in dark matter pair annihilation. Examples of models where the leptonic channels largely dominate include frameworks where either a discrete symmetry or the new physics mass spectrum suppresses other annihilation channels  \cite{Fox:2008kb,Harnik:2008uu}.
\item {\em Super-heavy dark matter models}: As pointed out in \cite{Cirelli:2008pk}, antiprotons can be suppressed below the PAMELA measured flux if the dark matter particle is heavy (i.e. in the multi-TeV mass range), and pair annihilates e.g. in weak interaction gauge bosons. Models with super-heavy dark matter can have the right thermal relic abundance, e.g. in the context of the minimal supersymmetric extension of the Standard Model, as shown \cite{Profumo:2005xd}.
\end{enumerate}
For those models the flux of antiprotons is generically suppressed to a level compatible with experimental data. 
For the three classes of models outlined above, we consider here the same large scale Galactic CR electron and positron spectrum adopted in Sec.\ref{sec:pulsars}
Both the pure $e^\pm$ model and in the lepto-philic models  allow a reasonable fit to both the PAMELA and the Fermi data is possible (though the latter seems to be 
favored). The preferred range for the dark matter mass lies between 400 GeV and 1-2 TeV, with larger masses increasingly constrained by the H.E.S.S. results \cite{hess,hess:09}. The required annihilation rates, when employing a conventional dark matter density profile (see \cite{interpretation_paper} for details), imply typical boost factors ranging between 20 and 100, when compared to the value $\langle\sigma v\rangle\sim3\times 10^{-26}\ {\rm cm}^3/{\rm sec}$ expected for a thermally produced dark matter particle relic. 
The super-heavy dark matter models are significantly disfavored by Fermi-LAT and H.E.S.S. CRE data. 
Notice that other dark matter models (including e.g. TeV-scale dark matter particles annihilating in muon-antimuon final states, either monochromatically or through the decays of intermediate particles) offer additional possible case-studies, as discussed {\it e.g.} in \cite{Bergstrominprep,Meade:2009iu}. 

%%----------------------------------------------------------------
%\begin{figure}[!t]
% %\begin{center}
% %\mbox{\epsfig{file=fermi.eps,width=0.5\textwidth}}\\[-1.0cm]
% %\end{center}
% \centering 
% \includegraphics[width=3in]{profumodifermi_smeared}
% \caption{Predictions for the CRE data from three benchmark dark matter models, and current measurements.
% The same large-scale Galactic CRE components (dotted line) as in Fig. \ref{fig:elepos_random_pulsars}
%  is used here. Statastical and systematic errors on Fermi-LAT data are added in quadrature. 
%  In the insert the positron fraction for the same models is compared with experimental data. 
%  For purely illustrative purposes, for each DM model we show both theoretical and smeared spectra obtained by accounting for a 15\% energy resolution.
% }
% \label{fig:fermi_DM}
%\end{figure}
%%----------------------------------------------------------------
%%----------------------------------------------------------------
%\begin{figure}[ht]
% %\begin{center}
% %\mbox{\epsfig{file=PAMELA.eps,width=0.5\textwidth}}\\[-1.0cm]
% %\end{center}
%  \centering 
% \includegraphics[width=4in]{profumodipamela}
% \caption{\footnotesize  \it The PAMELA positron-fraction data compared to the best fit models predictions in the dark matter interpretation.}
% \label{fig:PAMELA_DM}
%\end{figure}
%%----------------------------------------------------------------

\section{Conclusions}\label{sec:conclusions}

We reported on possible interpretations for the cosmic ray electron-plus-positron (CRE) spectrum measured by Fermi-LAT. The measured CRE flux is significantly harder than previously believed, and it does not show any sharp feature in the multi-hundred GeV range,  although there are hints of an extra-component between 100 GeV and 1 TeV. 

In the context of astrophysical interpretations to the CRE data, we discussed in the present analysis the case of a single large-scale diffuse Galactic (GCRE) component, and a two-component scenario which adds to the GCRE flux a primary electron and positron component  produced by mature pulsars.
In the GCRE scenario, a spatially continuous distribution of primary CRE sources in the Galactic disk, provides a satisfactory explanation to the Fermi-LAT CRE data for several 
combinations of the injection spectral index $\gamma_0$ and the CR propagation parameters.
This scenario, however, is in sharp tension with the PAMELA data on the positron fraction, more than previously 
considered in the framework of GCRE models, as a consequence of the hardness of the electron plus positron spectrum measured by Fermi-LAT.  
Furthermore,  a tension is also present between these GCRE models fitting the Fermi-LAT CRE spectrum and pre-Fermi experimental data below 10 
GeV and H.E.S.S. CRE data above the TeV. 
Taking into account nearby mature pulsars as additional sources of high-energy CRE, we showed that both the PAMELA positron excess and the Fermi CRE data are naturally explained by known objects. 

We also briefly considered another possible primary source of high-energy CRE: the annihilation or decay of particle dark matter in the Galactic halo. Fermi-LAT CRE data do not confirm the sharp spectral feature in the 500-1000 GeV range that prompted several studies to consider a dark matter particle mass in that same range. Yet, we showed that a dark matter particle annihilating or decaying dominantly in leptonic channels, and with a mass between 400 GeV and 2 TeV is compatible with both the positron excess reported by PAMELA and with the CRE spectrum measured by Fermi-LAT.  
%It is understood that the DM models considered here represent only a representative and limited subset of  of a much wider collection.  

While we found that the pulsar interpretation seems to be favored by Fermi-LAT CRE data, a clear discrimination between this and the dark matter 
scenario is not possible on the basis of the currently available data and may require to consider complementary observations.  Most relevant Fermi measurements in this framework will be:   
(i) extend the energy range both to lower and to higher energies than reported so far, (ii) allow anisotropy studies of the arrival direction of high-energy CRE, which could conclusively point towards one (or more than one) nearby mature pulsar as the origin of high-energy CRE, and (iii) deepen our understanding of pulsars via gamma-ray observations, and via the discovery of new gamma-ray pulsars, potentially extremely relevant as high-energy CRE sources.  Last but not least, Fermi measurements of the spectrum and angular distribution of the diffuse gamma-ray emission of the Galaxy will also shed light on the nature and spatial distribution of CRE sources. 

\section*{Acknowledgments}
The Fermi LAT Collaboration acknowledges support from a number of agencies and institutes for both development and the operation of the LAT as well as scientific data analysis. These include the NASA and the DOE United States, CEA/Irfu and IN2P3/CNRS in France, ASI and INFN in Italy, MEXT,
and the K. A. Wallenberg Foundation, the Swedish Research Council and the Swedish National Space Board in Sweden. Additional support from 
INAF in Italy for science analysis during the operations phase is also gratefully acknowledged. D.G. is supported by the Italian Space Agency under the contract AMS-02.ASI/AMS-02 n.I/035/07/0.  
%GALPROP development is funded via NASA grant NNX09AC15G.
%S.P. is partly supported by US DoE Contract DEFG02-04ER41268 and by NSF Grant PHY-0757911.


\begin{thebibliography}{99}

\bibitem{Kobayashi:2003kp}
 T.~Kobayashi, Y.~Komori, K.~Yoshida and J.~Nishimura,
 %``The most likely sources of high energy cosmic-ray electrons in  supernova
 %remnants,''
 ApJ\  {601}  (2004) 340.
% [arXiv:astro-ph/0308470].
 %%CITATION = ASJOA,601,340;%%

\bibitem{ams1} M.~Aguilar {\it et al.}, Phys.\ Reports {\bf 366} (2002) 331.

\bibitem{Strong:2007nh}
 A.~W.~Strong, I.~V.~Moskalenko and V.~S.~Ptuskin,
 %``Cosmic-ray propagation and interactions in the Galaxy,''
 {Ann.\ Rev.\ Nucl.\ Part.\ Sci.}\  {57} (2007) 285
% [arXiv:astro-ph/0701517]
 %%CITATION = ARNUA,57,285;%%
 
 \bibitem{Aharonian:95}
 F.~A.~Aharonian, A.~M.~Atoyan and H.~J.~V\"olk,\   
 A\&A\  294 (1995) L41.
 
 %\cite{Pohl:1998ug}
\bibitem{Pohl:1998ug}
  M.~Pohl and J.~A.~Esposito,
  %``Electron acceleration in SNR and diffuse gamma-rays above 1-GeV,''
   ApJ\  {507} (1998) 327.
  %arXiv:astro-ph/9806160.
  %%CITATION = ASTRO-PH/9806160;%%
 
\bibitem{atic} J.~ Chang {\it et al.} [ATIC Collaboration], Nature\ {456} (2008) 362. 

%%\cite{Torii:2008xu}
%\bibitem{Torii:2008xu}
%  S.~Torii {\it et al.}  [PPB-BETS Collaboration],
%  %``High-energy electron observations by PPB-BETS flight in Antarctica,''
%  arXiv:0809.0760 [astro-ph]. 
%  %%CITATION = ARXIV:0809.0760;%%

\bibitem{hess}
  F.~Aharonian {\it et al.}  [H.E.S.S. Collaboration],
  %``The energy spectrum of cosmic-ray electrons at TeV energies,''
  Phys.\ Rev.\ Lett.\  {\bf 101} (2008) 261104. 
  %[arXiv:0811.3894 [astro-ph]].
  %%CITATION = PRLTA,101,261104;%%

\bibitem{hess:09}
   F.~Aharonian {\it et al.}  [H.E.S.S. Collaboration],
  %``Probing the ATIC peak in the cosmic-ray electron spectrum with H.E.S.S,''
  arXiv:0905.0105 [astro-ph.HE].
  %%CITATION = ARXIV:0905.0105;%%

\bibitem{PAMELA}
 O.~Adriani {\it et al.}  [PAMELA Collaboration],\ 2009
 %``Observation of an anomalous positron abundance in the cosmic radiation,''
  Phys.\ Rev.\ Lett.\  {102} (2009) 051101.
 % arXiv:0810.4995 [astro-ph].  
 %%CITATION = ARXIV:0810.4995;%%

\bibitem{PAMELA_Nature}
O.~Adriani {\it et al.}  [PAMELA Collaboration],
Nature\  {\bf 458} (2009b) 607.

\bibitem{Fermi_CRE_1}
A.A.~Abdo et al. 2009 [Fermi collaboration],  Phys.\ Rev.\ Lett.\ 102,  (2009) 181101.
%arXiv:0905.0025.

\bibitem{Moskalenko:01} I.~ Moskalenko, A.~Strong, 
%``Models for Galactic cosmic-ray propagation,''
  Adv.\ Space Res.\  {\bf 27} (2001b) 717.
  %[arXiv:astro-ph/0101068].
  %%CITATION = ASRSD,27,717;%%  
  
\bibitem{interpretation_paper}
D.~Grasso {\it et al.}  [Fermi Collaboration],
  %``On possible interpretations of the high energy electron-positron spectrum
  %measured by the Fermi Large Area Telescope,''
  arXiv:0905.0636 [astro-ph.HE].
  %%CITATION = ARXIV:0905.0636;%%

\bibitem{Strong:04}
 A.W.~Strong,  I.W.~Moskalenko  and O.~Reimer,\ 
 %``Diffuse Galactic continuum gamma rays. A model compatible with EGRET data
 %and cosmic-ray measurements,''
ApJ\   613 (2004) 962.
 %[arXiv:astro-ph/0406254]
 %%CITATION = ASTRO-PH 0406254;%%

\bibitem{GeVexc}
A.A.~Abdo et al.~ [Fermi collaboration], 
%Fermi LAT measurements of the di?use gamma-ray emission at intermediate Galactic latitudes,
 submitted to Phys. Rev. Lett.  (2009b).

%\cite{DuVernois:2001bb}
\bibitem{DuVernois:2001bb}
  M.~A.~DuVernois {\it et al.} [HEAT Collaboration],
  %``Cosmic ray electrons and positrons from 1-GeV to 100-GeV: Measurements with
  %HEAT and their interpretation,''
  ApJ\ 559 (2001) 296.
  %%CITATION = ASJOA,559,296;%%

\bibitem{Harding:87}	
A.~K.~Harding and R.~Ramaty,\ 
%The Pulsar Contribution to Galactic Cosmic Ray Protons
Proc. of the 20th International Cosmic Ray Conference Moscow, Volume 2, (1987) 92.

%\bibitem{Chi:96}
%X.~Chi, K.~S.~Cheng and E.C.~Young,\   
%ApJ\  {459} (1996) 229.

\bibitem{Zhang:01}
L.~Zhang and K.S.~Cheng, A\&A\  368 (2001) 1063.

%\cite{Hooper:2008kg}
\bibitem{Hooper:2008kg}
 D.~Hooper, P.~Blasi and P.~D.~Serpico,
 %``Pulsars as the Sources of High Energy Cosmic Ray Positrons,''
 JCAP\  {\bf 0901} (2009) 025.
% [arXiv:0810.1527 [astro-ph]].
 %%CITATION = JCAPA,0901,025;%%

%\cite{Profumo:2008ms}
\bibitem{Profumo:2008ms}
 S.~Profumo,
 %``Dissecting PAMELA (and ATIC) with Occam's Razor: existing, well-known
 %Pulsars naturally account for the 'anomalous' Cosmic-Ray Electron and
 %Positron Data,''
 arXiv:0812.4457 [astro-ph].
 %%CITATION = ARXIV:0812.4457;%%

\bibitem{Manchester:05}
Manchester,~R.N., Hobbs,~G.B., Teoh,~A., \& Hobbs,~M.\ 2005, AJ, 129, 1993

\bibitem{Aharonian_book} 
F.A.~Aharonian,\  Very High Energy Cosmic Radiation, World Scientific, 2004.

\bibitem{blind_search}
A.A.~Abdo et al. [Fermi collaboration],  submitted to Science (2009c).

\bibitem{Pospelov:2007mp}
  M.~Pospelov, A.~Ritz and M.~B.~Voloshin,
  %``Secluded WIMP Dark Matter,''
  Phys.\ Lett.\  B {\bf 662} (2008) 53.
  %[arXiv:0711.4866 [hep-ph]].
  %%CITATION = PHLTA,B662,53;%%

\bibitem{ArkaniHamed:2008qn}
  N.~Arkani-Hamed, D.~P.~Finkbeiner, T.~Slatyer and N.~Weiner,
  %``A Theory of Dark Matter,''
  Phys.\ Rev.\  D {79} (2009) 015014.
  %[arXiv:0810.0713 [hep-ph]].
  
\bibitem{Fox:2008kb}
  P.~J.~Fox and E.~Poppitz,
  %``Leptophilic Dark Matter,''
  arXiv:0811.0399 [hep-ph].
  %%CITATION = ARXIV:0811.0399;%%

\bibitem{Harnik:2008uu}
  R.~Harnik and G.~D.~Kribs,
  %``An Effective Theory of Dirac Dark Matter,''
  arXiv:0810.5557 [hep-ph].

\bibitem{Cirelli:2008pk}
  M.~Cirelli  {\it et al.},
  %M.~Kadastik, M.~Raidal and A.~Strumia,
  %``Model-independent implications of the e+, e-, anti-proton cosmic ray
  %spectra on properties of Dark Matter,''
  Nucl.\ Phys.\  B {813} (2009) 1.
 % [arXiv:0809.2409 [hep-ph]].
 
 \bibitem{Profumo:2005xd}
  S.~Profumo,
  %``TeV gamma-rays and the largest masses and annihilation cross sections  of
  %neutralino dark matter,''
  Phys.\ Rev.\  D {\bf 72} (2005) 103521.
 %[arXiv:astro-ph/0508628].

%%%\cite{Adriani:2008zq}
%\bibitem{Adriani:2008zq}
%  O.~Adriani {\it et al.},\ 
%  %``A new measurement of the antiproton-to-proton flux ratio up to 100 GeV in
%  %the cosmic radiation,''
%  Phys.\ Rev.\ Lett.\  {\bf 102} (2009) 051101.
%  %[arXiv:0810.4994 [astro-ph]].
%  %%CITATION = PRLTA,102,051101;%%

\bibitem{Bergstrominprep}
 L.~Bergstrom, J.~Edsjo and G.~Zaharijas,
  %``Dark matter interpretation of recent electron and positron data,''
  arXiv:0905.0333 [astro-ph.HE].
  %%CITATION = ARXIV:0905.0333;%%

\bibitem{Meade:2009iu}
  P.~Meade, M.~Papucci, A.~Strumia and T.~Volansky,
  %``Dark Matter Interpretations of the Electron/Positron Excesses after
  %FERMI,''
  arXiv:0905.0480 [hep-ph].
  %%CITATION = ARXIV:0905.0480;%%

\end{thebibliography}
\end{document}